\documentclass[%
 twocolumn,
superscriptaddress,
amsmath,amssymb,
aps,
floatfix,
longbibliography,
]{revtex4}

\usepackage{graphicx}
\usepackage{dcolumn}
\usepackage{bm}
\usepackage{booktabs}
\usepackage{graphicx}
\usepackage{dcolumn}
\usepackage{bm}
\usepackage{nowidow}
\usepackage{tabularx}
\usepackage{siunitx}
\usepackage[usenames,dvipsnames]{xcolor}
\usepackage[colorlinks=true, allcolors=black]{hyperref}
\usepackage{tikz}
\usepackage{times}
\usepackage{hyperref}
\usepackage{afterpage}
\usepackage{float}
\hypersetup{
    citecolor = MidnightBlue,
    linkcolor = RedOrange
}
\usepackage{natbib}
\setcitestyle{square, comma, numbers,sort&compress, super}
\begin{document}

\newcommand{\YF}[1]{\textcolor{blue}{\,#1}}
\newcommand{\RS}[1]{\textcolor{magenta}{\,#1}}
\newcommand{\LI}[1]{\textcolor{olive}{\,#1}}
\newcommand{\ED}[1]{{\color{cyan}#1}}

\preprint{APS/123-QED}

\title{Controlling polymerization-induced phase separation in the synthesis of porous gels} 

\author{Yanxia Feng}
\affiliation{Department of Materials, ETH Z\"{u}rich, 8093 Z\"{u}rich, Switzerland.}%

\author{Noel Ringeisen}
\affiliation{Department of Materials, ETH Z\"{u}rich, 8093 Z\"{u}rich, Switzerland.}%

\author{Eric R. Dufresne}
\affiliation{Department of Materials Science \& Engineering, Cornell University, Ithaca, USA}%
\affiliation{Laboratory of Atomic and Solid State Physics, Cornell University, Ithaca, USA}%

\author{Lucio Isa}
\affiliation{Department of Materials, ETH Z\"{u}rich, 8093 Z\"{u}rich, Switzerland.}%

\author{Robert W. Style}
\email[]{robert.style@mat.ethz.ch}
\affiliation{Department of Materials, ETH Z\"{u}rich, 8093 Z\"{u}rich, Switzerland.}%

\date{\today}

\begin{abstract}
  Porous gels -- gels with solvent-filled pores that are much larger than their mesh size -- are widely used in engineering and biomedical applications due to their tunable mechanics, high water content, and selective permeability. 
Among various strategies to create porous gels, polymerization-induced phase separation (PIPS) has shown particular promise. 
However, the conditions that trigger and control PIPS remain poorly understood. 
Here, we systematically investigate the influence of solvent quality, polymeric precursor molecular weight, and polymer concentration on phase separation in polymerizing poly(ethylene glycol) diacrylate gels. 
We find that phase separation occurs when the precursor solution concentration 
is below the overlap concentration.
Phase-separated gels have a pore geometry that is controlled by solvent quality: better solvents result in smaller pores, while
worse solvents can create superporous, highly-absorbant gels.
Motivated by our results, we propose a theory that predicts when phase separation occurs in polymerizing gels, applicable across a wide range of polymer/solvent gel systems.
Our results provide a framework for the rational design of porous gels.

\end{abstract}

\maketitle


Gels -- and especially hydrogels -- are highly versatile materials with a wide range of applications. For example, the high water content of hydrogels makes them typically biocompatible, resembling the composition of most soft tissues in nature \cite{xue2021recent, grande1997evaluation}. Additionally, the ability of hydrogels to permit the diffusion of water-soluble molecules has led to their widespread use in areas including drug delivery \cite{bhattarai2010chitosan, hoffman2012hydrogels, tokarev2010stimuli}, catalysis \cite{diaz2011stimuli, zhang2014polymer}, and sensing technologies \cite{buenger2012hydrogels, zhou2024review, holtz1997polymerized}. Beyond these properties, gels can undergo significant volume changes, making them valuable as bio-inspired actuators in soft robotics and as absorbents \cite{lee2020hydrogel,kabiri2011superabsorbent,na2022hydrogel,zohuriaan2010advances}.

For many of these applications, it can be extremely beneficial to make porous hydrogels -- i.e. gels with solvent-filled pores that are larger than the gel's mesh size.
This is because adding such porosity can dramatically change the gel's permeability, swelling kinetics, and the diffusivity of solutes and particulates through the gel \cite{yazdi2020hydrogel, wu2010water, eddine2023sieving, eddine2024tuning,alsaid2021tunable}.
For example, tissue scaffolds require \emph{macroporous} hydrogels to allow cell growth and nutrient transport \cite{drury2003hydrogels,zhu2011design,el2013hydrogel}.
Hydrogels for drug-delivery need tailored pore sizes to control drug release rates \cite{li2016designing,liu2024advances,hamidi2008hydrogel}.
Porous hydrogel membranes need controlled pore sizes to achieve selective filtration capabilities (e.g. for use in wastewater treatment) \cite{radoor2024recent,sinha2019advances}.
Swelling applications, such as spill absorption or actuation, require gels with very large pores, such as superporous hydrogels \cite{chen1999synthesis,kuang2011polysaccharide,moser2022hydroelastomers}.

Given the importance of porosity in gel performance, we need reliable techniques for accurately introducing porosity.
Broadly speaking, these can be broken down into four groups:
First, in \emph{Templating}, gels are formed around porogens (e.g. salt, ice, emulsion droplets or gas bubbles) that are later removed \cite{annabi2010controlling,nam2000novel,eiselt2000porous,liang2021anisotropically,joukhdar2021ice,zhang2019emulsion}.
Second, in \emph{Granular Assembly}, gel particles are jammed or bonded together \cite{hirsch20213d,puiggali2023growth,highley2015direct}.
In \emph{Light-Based Structuring}, well-defined light patterns are used to photopolymerize (or degrade) gels to create pores \cite{xing2015two,mondschein2017polymer}.
Finally, in \emph{Polymerization-Induced Phase Separation} (PIPS), polymerizing solutions spontaneously phase-separate to form gels with solvent-filled pore networks \cite{ishikawa2023percolation,wu2010water, danielsen2023phase}.
Out of these techniques, PIPS is potentially the most powerful option.
This is because it can make large volumes of materials, it does not require long porogen leaching steps, and the resulting materials have good mechanical strength \cite{foudazi2023porous,nicol2021photopolymerized,zhang2021stretchable}.
Furthermore, it has been shown to be capable of producing hydrogels with a range of pore sizes from 10s of nm to 10s of $\mu$m \cite{nicol2021photopolymerized}.
However, PIPS requires extensive empirical testing to determine the conditions necessary for making gels with given pore sizes, or even to determine when a polymerizing system will phase separate during gelation \cite{foudazi2023porous}.
To overcome this issue, we need a quantitative understanding of the fundamental processes that drive PIPS.

Here, we shed light on the key processes that underpin porous-gel formation via PIPS.
We create gels formed from polymer/solvent combinations with a range of different solvents.
Gels phase separate during polymerization to form porous gels when the polymer content in the gel, $\phi_p$, is below a critical value, $\phi_{ps}$. 
We show that $\phi_{ps}$ is essentially equal to the overlap concentration, $\phi^*$, of precursor in the solvent, and rationalize this observation by considering the microscopic processes involved in polymerization.
Furthermore, we show how polymer/solvent compatibility affects the resulting pore structure.
Our findings provide design principles for selecting polymer molecular weight, solvent type, and polymer concentration to fabricate porous gels via PIPS.

\section*{Results and Discussion}
\subsection*{Phase separation in PEGDA gels\label{sec:phenomena}}

We study porous gel synthesis via PIPS by using polyethylene glycol diacrylate (PEGDA) oligomers.
These are short, hydrophilic, polyethene glycol (PEG) chains, that are capped at each end with acrylate groups.
Thus, each oligomer can bond with up to four other PEGDA molecules when polymerized, and so PEGDA cross-links into a polymer network upon the initiation with a photoinitiator and UV light \cite{li2023vitro,feng2025characterizing}. 
In general, PEGDA is polymerized in the presence of a solvent (e.g. water) to form a gel.
As this process is particularly simple, PEGDA is widely used to make hydrogels \cite{browning2014determination,testa2023surface,yang2024dehydration,lin2009peg,wu2010water}. 

\begin{figure}[h]
    \centering
    \includegraphics[width=1.0\columnwidth]{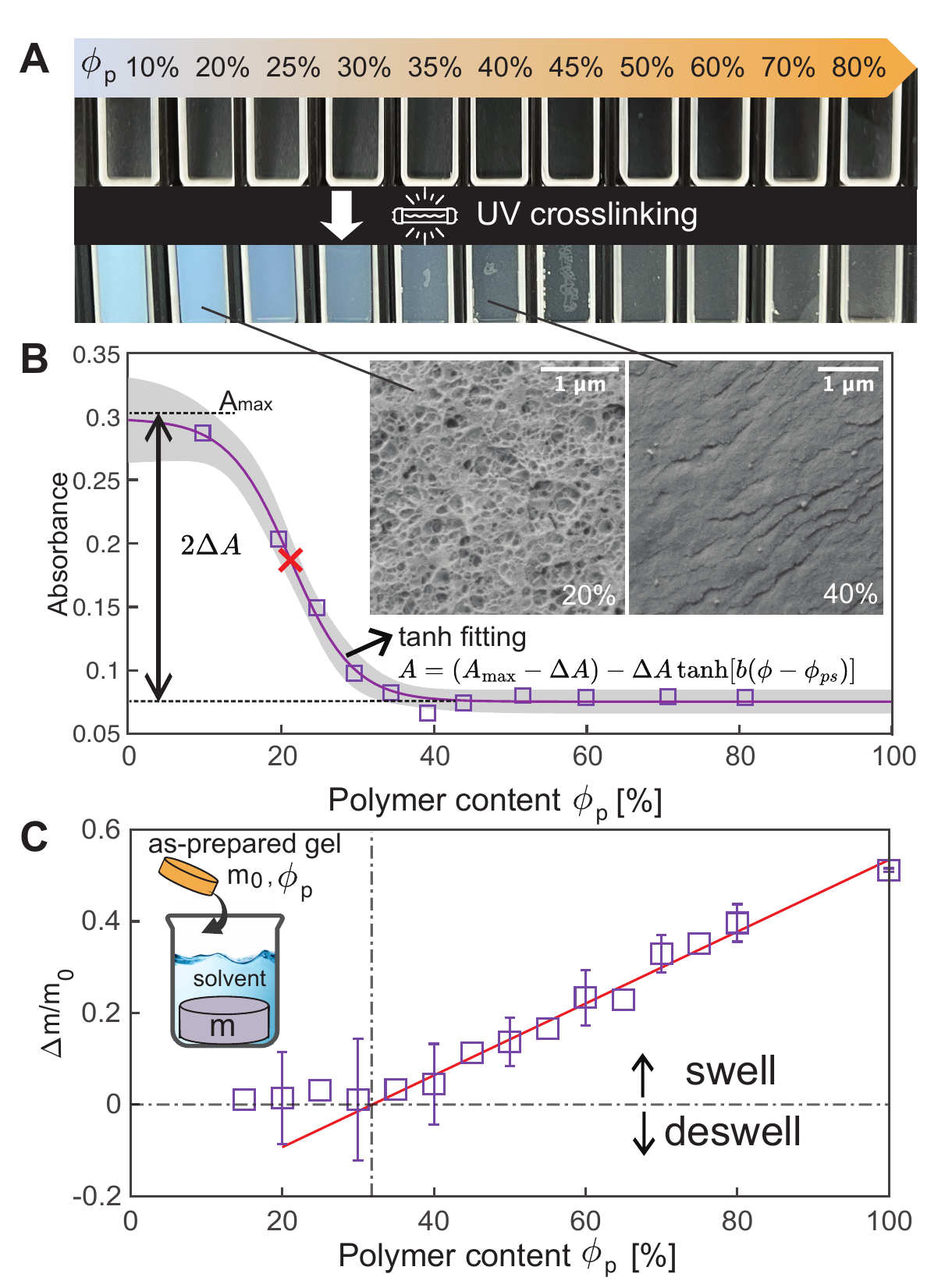}
    \caption{Phase separation in PEGDA hydrogels. 
    \textbf{A.} PEGDA 700/water precursor solutions (top) and the resulting hydrogels (bottom). The polymer content increases from left to right as marked. PEGDA is fully miscible in water, so all precursor solutions are transparent. After cross-linking, low $\phi_p$ samples become opaque (samples are 2 mm  thick).
    \textbf{B.} 
    Sample opacity, as characterized by the absorbance of 600-nm light passing through the samples in (A). 
    The continuous curve is a best fit sigmoidal curve.
    The center of this fit gives a measure of the phase separation boundary, $\phi_{ps}$.
    \textbf{Inset: }Cryo-EM images of hydrogels with $\phi_p=$20,40\%. The 20\% sample is porous with $\sim$100~nm, polydisperse pores, indicating the presence of phase separation. By contrast, the 40\% sample appears homogeneous at the same scale.
    \textbf{C.} The relative swelling of as-prepared hydrogels when immersed in water. 
    Low $\phi_p$ samples (left of the vertical line) do not swell. Higher $\phi_p$ samples do swell. A comparison with (B) shows that only the transparent samples swell.}
    \label{fig:uvdata}
\end{figure}

When PEGDA solutions with different polymer contents are cross-linked to form gels, there are stark differences in the turbidity of the resulting gels (Figure \ref{fig:uvdata}A).
We prepare hydrogels from solutions of PEGDA 700 (700 g/mol, 13 repeat PEG units) in de-ionized water, with different polymer volume percentages, $\phi_p$.
The gels are prepared via UV polymerization using 2-hydroxy-2-methylpropiophenone as an initiator (see the Supporting Information for a discussion on how the results are not affected by changes in the initiator chemistry). 
Before polymerization (top row), all PEGDA solutions are transparent, indicating complete miscibility of PEGDA and water.
After polymerization (bottom row), all the solutions form hydrogels. Gels with low $\phi_p$ become visibly opaque, while those with $\phi_p\gtrsim 40\%$ remain transparent.

The increased gel turbidity at low $\phi_p$ is caused by the formation of a swollen, porous gel containing solvent-filled pores \cite{wu2010water}.
We visualize these pores by performing cryo-electron microscopy on sections of the gels with $\phi_p=20,40\%$ from Figure \ref{fig:uvdata}A (see insets, Figure \ref{fig:uvdata}B).
The 20\% gel clearly has a porous structure, while we see no porous structure in the 40\% gel.
In the former case, the $O( 100$ nm$)$ pores are large enough to scatter light. 
Thus, their presence explains the sample opacity.

We characterize the changes in sample turbidity -- and thus the onset of porous gel formation -- by measuring the absorbance, $A$, of light as it passes through 2 mm-thick samples at 600~nm (e.g. \cite{wu2010water, van2021generic, raymond2022spectrophotometric}). 
Absorbance is calculated from the fraction of transmitted light, $T$, as $A = -\log_{10}(T)$, and represents the total loss of light due to both absorption and scattering.
The results are given in Figure \ref{fig:uvdata}B, clearly showing the transition from turbid to transparent as $\phi_p$ increases.
This data is well described by a sigmoidal tanh function 
\begin{equation}
    A=(A_{\mathrm{max}}-\Delta A)-\Delta A\tanh[b(\phi_p-\phi_{ps})],
\end{equation}
as shown by the continuous curve in the Figure (see the Supporting Information for fitting parameters values). Here, $2\Delta A$ gives the difference in absorbance between gels with a fully developed porous structure (low $\phi_p$) and homogeneous gels (high $\phi_p$).
The particular use of this fit is that it allows us to extract the polymer concentration below which the gel switches from transparent to opaque -- here, $\phi_{ps}=21.1 \pm 1.5 \%$. 
In fact,  this transition is a marker for the onset of phase separation during polymerization.

To confirm that changes in sample turbidity are caused by phase separation, we measure how as-prepared gels swell in bulk water.
We place freshly-prepared  PEGDA gels into pure water to allow them to reach swelling equilibrium (for at least 5 hours).
Then, we measure their relative change in mass, $\Delta m/m_0$, where $\Delta m$ is the change in mass and $m_0$ is the as-prepared sample mass.
The results are shown in Figure  \ref{fig:uvdata}C.
Gels with $\phi_p
<$ 30\% do not swell at all in water.
Gels with $\phi_p>$ 30\% draw in significant amounts of water.
This transition from swelling to non-swelling closely matches the transition from turbid to transparent gels \cite{wu2010water}.
This match is explained by the low-$\phi_p$, turbid gels being phase-separated: these gels have bulk water in their pores, and are thus are in equilibrium with bulk water -- adding further bulk water will not change the swelling equilibrium.
By contrast, the high-$\phi_p$, transparent gels swell in water.
Thus, they are not in equilibrium with bulk water in their as-prepared state, indicating that they do not contain large, water-filled pores (\emph{i.e.} they are not phase separated).
Taken together, the results above confirm that we can use optical measurements of gel turbidity to establish when phase separation occurs.
Furthermore, we can use  $\phi_{ps}$ as a convenient measure of the critical polymer content, below which phase separation occurs.

This type of phase separation is a general phenomenon that is not limited to the specific PEGDA/water combination.
We show this by preparing gels from PEGDA and a range of solvents with varying polarities.
In particular, we make gels from solutions of PEGDA 700 in acetonitrile, acetone, ethanol, 2-butanol, and isopropanol.
Furthermore, we also make gels with PEGDA 575 (575 g/mol, 10 PEG monomers per chain).
In all cases, we see similar phase-separation behavior to that in Figure \ref{fig:uvdata}A:
low-$\phi_p$ gels phase separate, while high-$\phi_p$ gels are homogeneous and transparent (see pictures in the Supporting Information).
For each combination, the phase-separation boundary, $\phi_{ps}$ (measured optically, as above) is given in Figure \ref{fig:overlapagainsps}A.

Comparing $\phi_{ps}$ for different polymer/solvent combinations sheds light on the dominant factors that control when phase separation occurs.
In Figure \ref{fig:overlapagainsps}A, we see that $\phi_{ps}$ can vary by up to $25\%$ for different solvents.
In particular, more hydrophobic solvents that are expected to be worse solvents for PEGDA (2-butanol and isopropanol) have high $\phi_{ps}$, while the better, more hydrophilic solvents (water and acetonitrile) have low $\phi_{ps}$.
Furthermore, gels made from the shorter PEGDA 575 have a higher $\phi_{ps}$ than gels made from PEGDA 700.
This implies that phase separation happens more easily for worse quality solvents for PEGDA, and for shorter oligomers.

\subsection*{Phase separation is determined by the overlap concentration}

To more quantitatively test the link between solvent quality and phase separation, we measure the volume percentage at which oligomers overlap in solution, $\phi^*$.
This characterizes solvent quality as it is inversely proportional to the swollen volume of an oligomer in a dilute solution, $V_{sw}$ ($\phi^* = V_p/V_{sw}$, where $V_p$ is the volume of a dry oligomer).
Good solvents cause oligomers to be highly swollen, while bad solvents cause them to collapse to a low $V_{sw}$ -- \emph{i.e.} better solvents have lower $\phi^*$.
We measure $\phi^*$ via
microrheology for all the polymer/solvent combinations except those involving acetonitrile and acetone (these dissolved the fluorescent tracer particles that we require for the technique).
\begin{figure*}[htpb]
    \centering
    \includegraphics[width=0.8\linewidth]{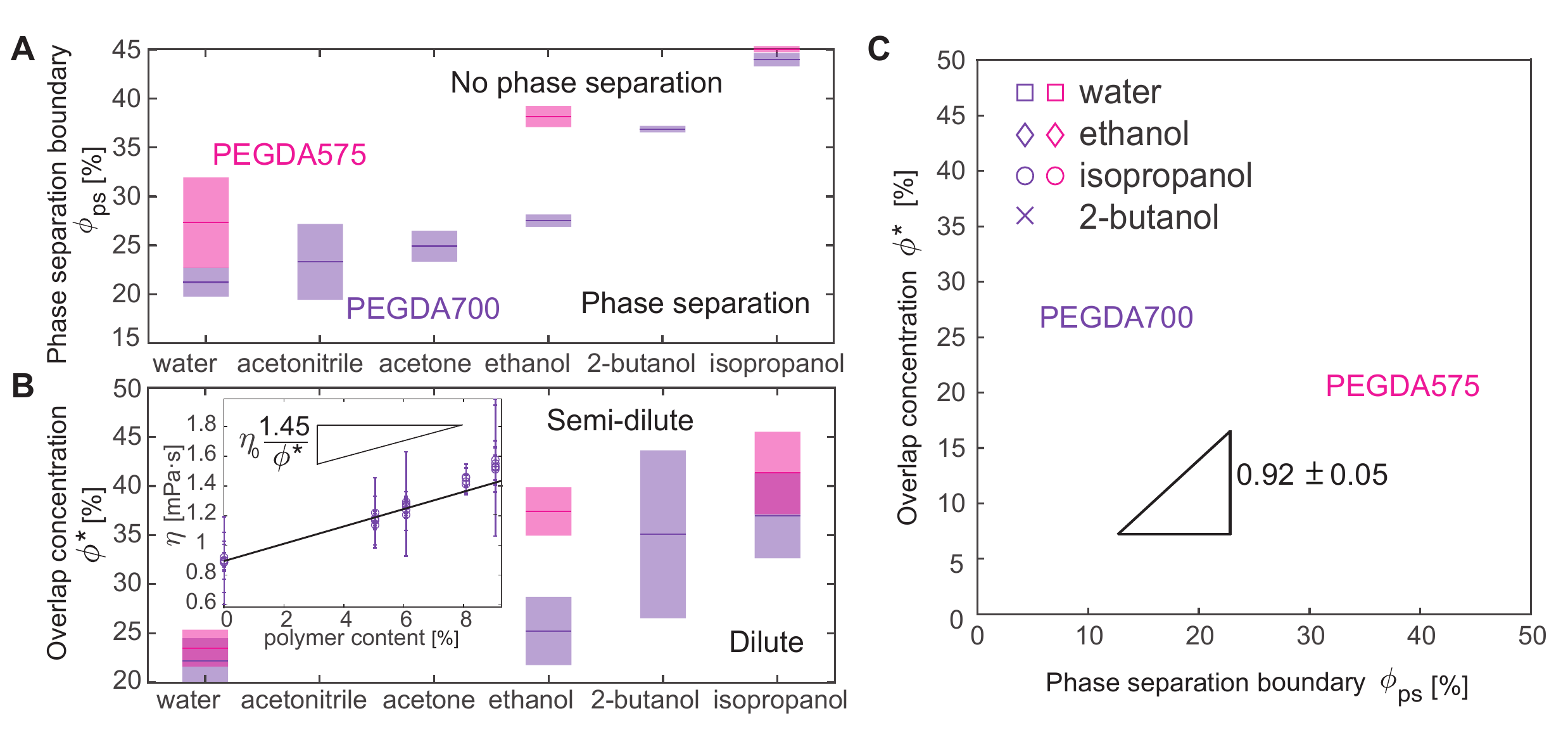}
    \caption{The effect of changing solvent on phase separation during gel formation. 
    \textbf{(A)} The critical concentration, $\phi_{ps}$, at which phase separation occurs for different solvents and PEGDA chain lengths.        
    \textbf{(B)} The overlap concentration, $\phi^*$, for different PEGDA/solvent combinations. Inset: we calculate $\phi^*$ from microrheology measurements of viscosity in dilute solutions. Here, the results are for PEGDA700/water.
    \textbf{(C)} A comparison of $\phi_{ps}$ and $\phi^*$ for different polymer–solvent combinations reveals an excellent correlation. The black line is a linear fit.
    }
    \label{fig:overlapagainsps}
\end{figure*}

We extract $\phi^*$ from measurements of the shear viscosity, $\eta$, of dilute solutions of different concentrations.
In this dilute limit, $\eta$ is a linear function of $\phi$ (e.g. inset Figure \ref{fig:overlapagainsps}B).
Thus, we can fit the $\eta(\phi)$ for each polymer/solvent combination with the relationship $\eta=\eta_0(1 + 1.45\phi/\phi^*$), where $\eta_0$ is the viscosity of the pure solvent \cite{tirtaatmadja2006drop,rodrigues2020critical} (see Methods and Supporting Information).
\emph{i.e.} we obtain $\phi^*$ from the slope of $\eta(\phi)$, as shown in the Figure.

Figure \ref{fig:overlapagainsps}B shows the resulting values of $\phi^*$.
As expected, hydrophilic solvents like water are good solvents for PEGDA (low $\phi^*$), while more hydrophobic solvents like isopropanol are worse solvents for PEGDA (higher $\phi^*$).
Additionally, a qualitative comparison of $\phi^*$ and $\phi_{ps}$ in Figures \ref{fig:overlapagainsps}A,B confirms our initial impression that good solvents lead to less phase separation (lower $\phi_{ps}$).

In fact, there is a strong quantitative correlation between $\phi^*$ and the onset of phase separation for all our samples.
In Figure \ref{fig:overlapagainsps}C, we plot $\phi_{ps}$ and $\phi^*$ against each other.
There is a very good, linear correlation, $\phi^*=\alpha \phi_{ps}$, with
$\alpha=0.92 \pm 0.05$ being very close to unity.
\emph{i.e.} $\phi^*$ and $\phi_{ps}$ take essentially the same values.

\begin{figure}[htbp]
    \centering
    \includegraphics[width=0.9\linewidth]{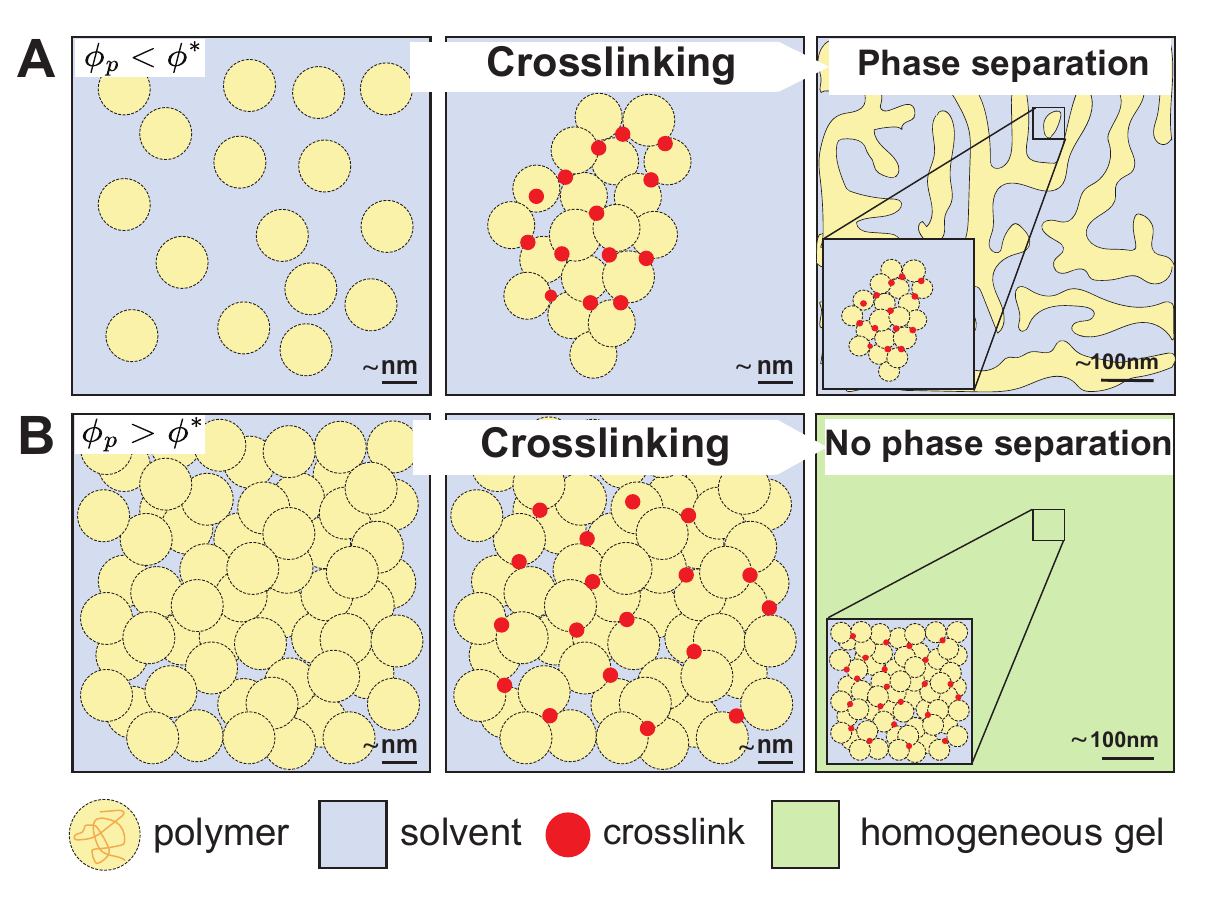}
    \caption{ Phase separation occurs below the overlap concentration. A) Below the overlap concentration: $\phi_p<\phi^*$. 
    When oligomers polymerize, they form multiple connections to neighboring oligomers. These connections cause oligomers to clump together, resulting in dense gel regions. The local densification leaves behind regions that are depleted of polymer.
    Thus the final structure is a phase-separated mixture of dense gel and solvent-filled pores.
    B) Above the overlap concentration: $\phi_p>\phi^*$. The oligomers can bond into networks without local densifying. Thus, the final structure is not phase separated.}
\label{fig:mechanism}
\end{figure}
These results suggests that $\phi^*$ controls when phase separation occurs.
This makes sense, if we think about the microscopic gelation process (Figure \ref{fig:mechanism}).
When $\phi_p<\phi^*$, oligomers in the precursor solution are spaced apart from each other.
If there is a driving force causing the oligomers to phase separate as they polymerize -- as here (see below) -- 
then oligomers will cluster as they form a gel. 
This will leave liquid-filled pores in the resulting gel (Figure \ref{fig:mechanism}A).
By contrast, when $\phi_p>\phi^*$, the oligomers in the precursor solution already overlap, and are thus compressed relative to their equilibrium swelling in excess solvent \cite{flory1942thermodynamics}.
They can therefore polymerize without significantly moving, and thus the gel will form homogeneous, dense blocks of compressed chains (see Figure \ref{fig:mechanism}B).
Such a gel should also swell when placed in pure solvent, as the now-crosslinked chains swell towards their equilibrium swelling configuration (see Figure \ref{fig:uvdata}C).
Thus, we propose that $\phi^*$ of an oligomer in a polymer precursor solution directly determines whether the resulting gel will phase separate or not upon crosslinking.

\subsection*{The mechanism for phase separation}

A key remaining question is why the oligomers clump together (\emph{i.e.} phase separate) as they polymerize \cite{zwicker2022intertwined}.
If they did not do this, we would simply obtain a gel that was a homogeneous mesh of oligomers, without observable phase separation. 
Thus, we need a driving phase-separation mechanism -- known in the literature as a \emph{syneresis} mechanism -- which can take one of two different forms \cite{okay2000macroporous,duvsek2006structure}.
First, \emph{solvent-induced syneresis} is caused by somewhat poor polymer/solvent compatibility, where short molecules of the polymer backbone in the gel are soluble in the solvent (due to their large entropy of mixing), but long molecules are not.
Thus, growing polymer gel molecules will phase separate when they surpass a critical molecular weight.
Second, \emph{cross-linking-induced syneresis} is caused by the addition of cross-linking points to the polymerizing gel network.
The more cross-linked the network is, the denser it becomes.
When enough cross-linking points are added to a polymerizing solution, the resulting gel can become so dense that it cannot fill the whole original volume.
Then, the gel will phase separate into separate phases of dense gel and pure solvent \cite{ishikawa2023percolation}.
Note that this mechanism also works in good solvents, where solvent-induced syneresis does not work.

Here, both mechanisms driving PIPS are important, with the more hydrophobic solvents undergoing solvent-induced syneresis, while the more hydrophilic ones undergo cross-linking-induced syneresis.
As a quick test of this, we examine the solubility of long-chain (12 kDa), linear PEG molecules in the different solvents (see Supplement for further information).
PEG is very similar to polymerized PEGDA -- so if the PEG is insoluble in a solvent, we expect solvent-induced syneresis to occur during polymerisation (see further details in the Supplement).
Figure \ref{fig:solubility}A shows pictures of 10 vol\% solutions in front of a background with a black cross.
PEG solutions in water, acetonitrile and acetone are transparent, implying that the PEG is soluble in these solvents.
PEG solutions in ethanol, 2-butanol and isopropanol are opaque, as the polymer is insoluble.
Thus, PEGDA can undergo solvent-induced phase separation in these last three solvents.
By contrast, cross-linking-induced syneresis likely dominates for PEGDA in water, acetone and acetonitrile (see Supporting Information for further evidence).

\begin{figure}[htbp]
    \centering
    \includegraphics[width=\columnwidth]{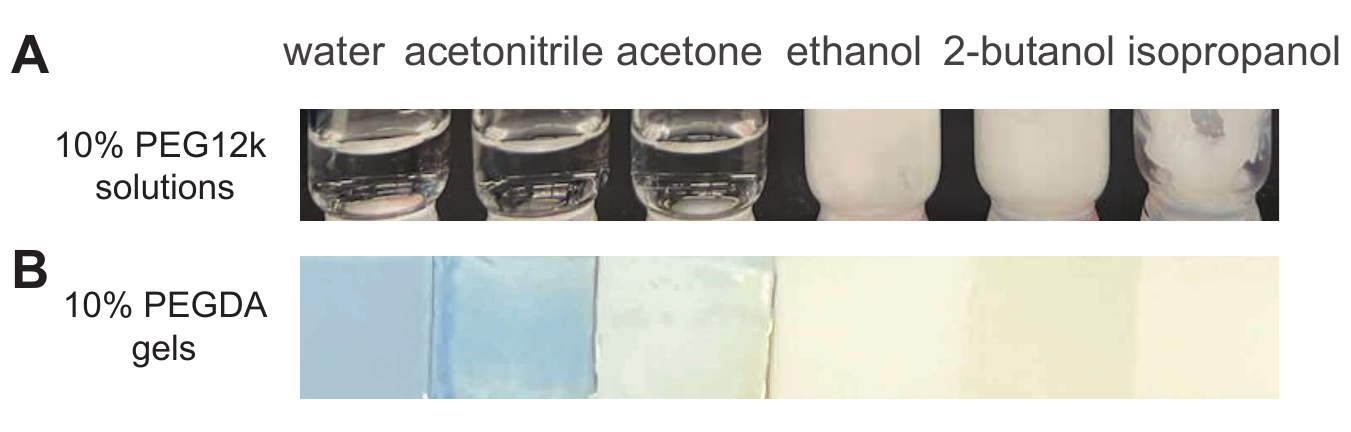}
    \caption{Distinguishing between solvent-induced and cross-linking-induced syneresis mechanisms based on PEG solubility and resulting gel opacity.
    \textbf{(A)} Visual assessment of 10 vol\% solutions of linear 12 kDa PEG in various solvents, placed in front of a black background to indicate solution clarity. Transparency in water, acetonitrile, and acetone suggests PEG solubility, whereas opacity in ethanol, 2-butanol, and isopropanol suggests poor solubility.
    \textbf{(B)} Corresponding appearance of PEGDA gels polymerized in these solvents at 10 vol\% of PEGDA700. Gels formed in PEG-soluble solvents are semi-transparent and bluish, whereas those formed in PEG-insoluble solvents are fully opaque. 
    }
    \label{fig:solubility}
\end{figure}

While the particular mechanism driving phase separation does not appear to change the relationship between $\phi^*$ and $\phi_{ps}$ (see results for different polymer/solvent combinations in Figure \ref{fig:overlapagainsps}C), it should control a gel's final pore structure.
Solvent-induced syneresis is known to produce gels with larger, more random pores, while cross-linking-induced syneresis produces gels with smaller, more uniform pores \cite{okay2000macroporous}.
This matches our experimental observations.
In particular, the solvent-induced syneresis gels (ethanol, 2-butanol and isopropanol) are completely opaque (Figure \ref{fig:solubility}B), suggesting that there is significant light scattering off pores, and that the pore size is comparable to or bigger than the wavelength of light (\emph{i.e.} $O(500$ nm$)$).
By contrast, the cross-linking-induced syneresis gels (water, acetonitrile and acetone) are blueish, and hazy instead of opaque.
This indicates Rayleigh scattering of light in gels, off pores that are significantly smaller than the wavelength of light.
Overall, this implies that we can tailor pore sizes by controlling solvent quality, with better solvents leading to smaller pores.

\subsection*{Fabricating superporous gels}

In fact, the dependence of pore size on solvent quality enables the synthesis of superporous hydrogels \cite{omidian2005advances}.
Most conventional hydrogels shrink considerably upon drying, ultimately forming a dense, solid polymer block. 
However, if the gel contains sufficiently large pores, air can invade these voids during drying, resulting in a sponge-like, air-filled structure \cite{tsotsas2011modern}. 
These superporous gels absorb solvent much faster than conventional hydrogels because capillary action rapidly draws solvent into the pore network, followed by short-range diffusion that swells the polymer matrix.

\begin{figure}[htbp]
    \centering
    \includegraphics[width=\columnwidth]{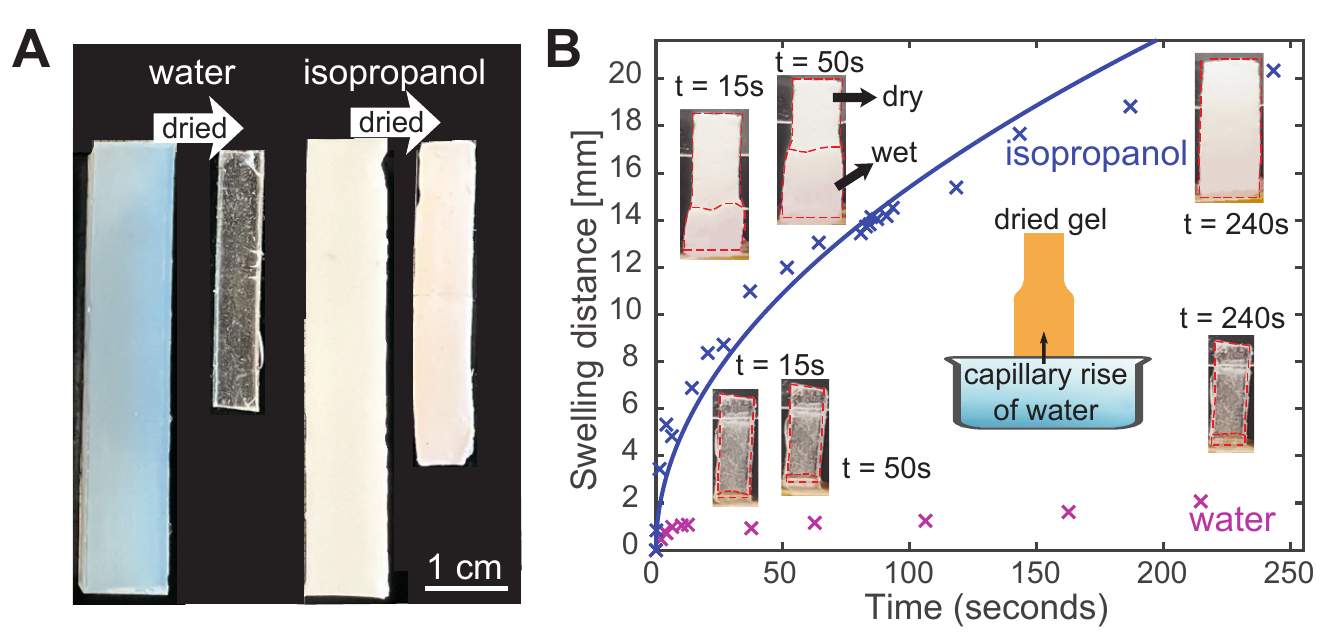}
    \caption{Comparision of drying and swelling behaviors of superporous PEGDA gel and conventional PEGDA hydrogel.
    \textbf{(A)} Comparison of dried PEGDA700 gels ($\phi_p = 20\%$) prepared in water versus isopropanol. The water-based gel undergoes significant shrinkage during drying, reducing to $\sim$60\% of its original width, while the isopropanol-based gel retains $\sim$80\% of its original width, indicating the presence of air-filled pores and a superporous structure.
    \textbf{(B)} Swelling height as a function of time when dried samples are in contact with water. Inset pictures show time-lapse of water absorption by dried gels. The superporous gel exhibits rapid capillary uptake, with the swelling front reaching $\sim$25\% of the sample height in 10 seconds, and full swelling in under 4 minutes. In contrast, the conventional water-based gel swells very slowly, with the swelling front barely advancing after 20 minutes. The continuous curve shows a best fit of Washburn's law ($h\propto t^{1/2}$).
    }
    \label{fig:superporous}
\end{figure}

We demonstrate how to make superporous PEGDA-based gels by using a poor solvent like isopropanol.
We prepare two precursor solutions of PEGDA 700 at a polymer concentration of $\phi_p = 20\%$ in water and isopropanol, then cure the mixture under UV light.
The resulting gels are dried under ambient conditions until the polymer content exceeds approximately 85\%. 
As shown in Figure~\ref{fig:superporous}A, a PEGDA700 gel formed in water is hazy and blue-colored, and stiffens and shrinks significantly during drying, with the final width reduced to about 60\% of the original length: roughly what we expect for a hydrogel dried to a polymer content of 85\% with no air invasion of pores during drying.
By contrast, a PEGDA700 gel formed in isopropanol is opaque, and shrinks significantly less upon drying than the water-derived gel.
Furthermore, the dried gel is soft and spongy, indicating that it has air-filled pores, and thus is superporous \cite{omidian2005advances}.

To confirm that the isopropanol-derived gel is superporous, we clamp a $\sim$2 cm tall sample and wet its base with water.
As shown in Figure~\ref{fig:superporous}B, the gel rapidly absorbs water, taking only about four minutes for water to completely swell the full height of the sample. 
This fast wetting flow is driven by capillary wicking into the air-filled pores of the dry gel.
As such, the position of the wetting front can be described by Washburn's law for capillary rise height in a hydrophilic porous medium: $h=(\gamma_l r_pt/2\eta_l)^{1/2}$, where $r_p$ is the average pore radius, and $\gamma_l$ and $\eta_l$ are the surface tension and dynamic viscosity of the liquid, respectively.
We fit Washburn's law to the data in Figure \ref{fig:superporous}B to obtain an average dry pore radius of $65$ nm.
By comparison, the water-derived PEGDA gel exhibited only minimal water swelling: even after 20 minutes, the swelling front remained barely visible.
This confirms the superior absorption kinetics of the superporous variant.

\subsection*{Implications for porous gel synthesis}
\begin{figure}[h]
    \centering
    \includegraphics[width=\columnwidth]{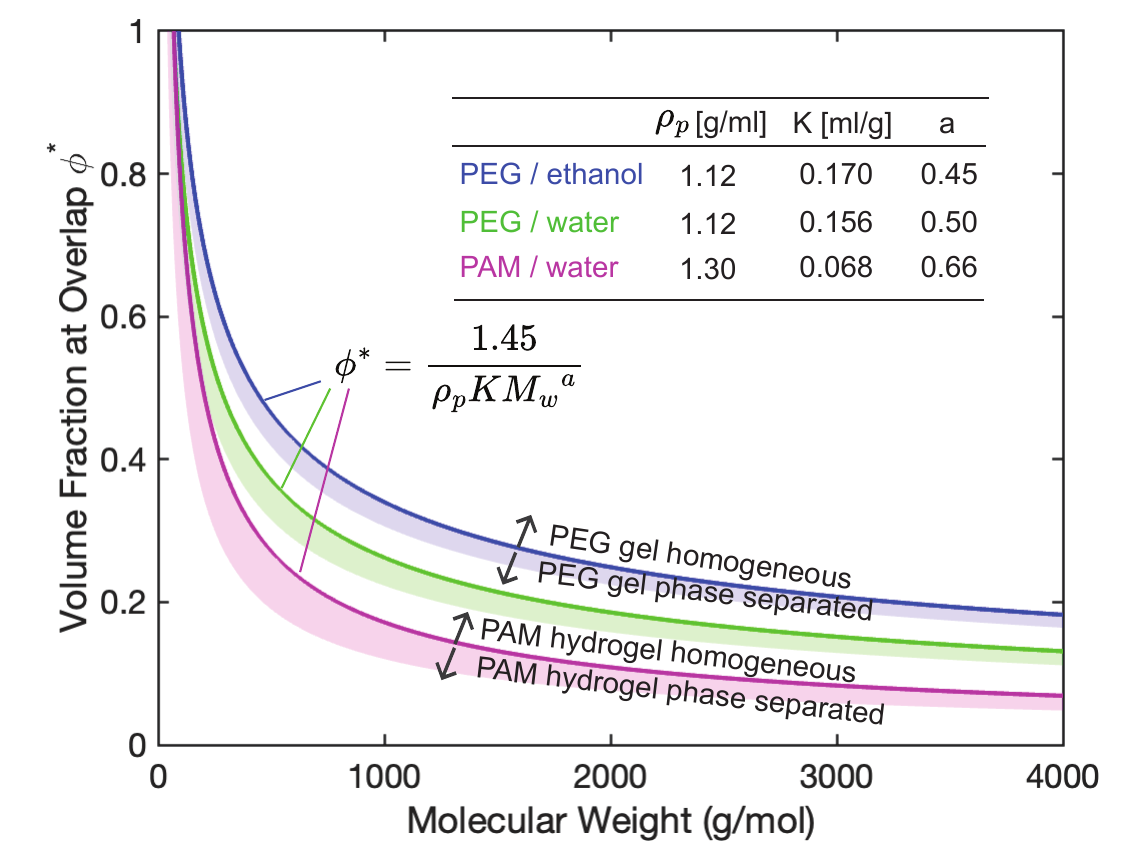}
    \caption{Predicted overlap concentration $\phi^*$ versus molecular weight, $M_w$, for PEG/water, PEG/ethanol and polyacrylamide/water solutions. We predict that porous gels will result when gels are made from polymeric precursors with molecular weight $M_w$ in a solution with $\phi_p<\phi^*$.}
    \label{fig:phi_star}
\end{figure}

Our findings offer insights into the design of phase-separating gel systems.
For example, they allow us to predict the onset of phase separation for a system where oligomers of a given molecular weight, $M_w$, are end-crosslinked together in a solvent to form a gel.
In particular, we can predict $\phi^*$ for such a system by combining the result that $\phi^* = 1.45/[\eta]$ \cite{rodrigues2020critical} with the Mark-Houwink equation for the intrinsic viscosity of an oligomer solution \cite{teraoka2002introduction}:
$[\eta] = \rho_p K {M_w}^a$, 
Here, $\rho_p$ is the density of the polymer, while $a$ and $K$ are constants that are fixed for a given polymer/solvent combination, and which can be found in standard literature references (e.g. \cite{dohmen2008hydrodynamic,brandrup1999polymer}). 
Combined, these equations yield the condition for the onset of PIPS:
\begin{equation}
    \phi_{ps}\approx\phi^*=\frac{1.45}{\rho_pK {M_w}^a}.
    \label{eq:KM}
\end{equation}
 Figure \ref{fig:phi_star} shows typical predictions of PIPS onset.
 The figure shows $\phi^*$ as a function of the molecular weight of an oligomer for three different systems: PEG-water, PEG-ethanol and polyacrylamide (PAM)-water.
 For each system, the parameters $a$ and $K$ are taken from \cite{dohmen2008hydrodynamic,brandrup1999polymer} and are listed in the Figure.
 Our results suggest that PIPS should only occur below these curves, and the predictions match the features of our experiments.
All the curves decrease rapidly with increasing $M_w$, showing that it is  easier to form phase-separated gels with shorter oligomers than longer ones -- agreeing with the fact that PEGDA 575 gels phase separate at higher $\phi$ than PEGDA 700 gels (Figure \ref{fig:overlapagainsps}).
Furthermore, the curve for PEG-ethanol mixtures lies higher than the one for PEG-water mixtures -- matching our experimental observations that PEGDA gels in ethanol phase separate at higher concentrations than PEGDA gels in water.

Here, we have focused on the end cross-linking of oligomers to form gels.
However, the ideas presented in Figure \ref{fig:phi_star} may also give insight into PIPS in gels formed from solutions of monomers and crosslinkers.
A common example is PAM/water (purple curve, Figure \ref{fig:phi_star}).
These gels are known to phase separate at certain polymer concentrations and monomer/crosslinker ratios \cite{gombert2020hierarchical}.
In gels like these, we take a simplified view of the polymerization process, whereby the monomers (molecular weight $M_m$) first polymerize to form oligomer chains, and then these are connected into a network by the cross-linking molecules.
Assuming a perfect network, the average number of monomers in each oligomer should then be $N_o=2N_m/(f N_c)$, where $N_m$ is the total number of monomers, $N_c$ is the total number of crosslinkers, and $f$ is the functionality of the crosslinkers.
Thus, the average molecular weight of `oligomers' between crosslinks is 
$M_o=2 M_m N_m/(f N_c)$, and we can use this value as the molecular weight in equation (\ref{eq:KM}) to predict PIPS (purple curve, Figure \ref{fig:phi_star}).
The chief implication is that the smaller $M_o$ is, the easier it is to phase separate.
Indeed, this agrees with previous experimental work in PAM gels, which has shown that increasing crosslinker concentration (i.e., increasing $N_c$ and reducing $M_o$) leads to greater opacity -- and thus more phase separation \cite{gombert2020hierarchical}.

\section*{Conclusions}

In conclusion, we have identified the conditions necessary to form porous gels  via polymerization-induced phase separation of functional oligomers dissolved in a solvent. 
Phase separation occurs whenever the oligomer concentration in the solvent is below the overlap concentration.
The characteristics of the resulting pores in the gel are strongly affected by solvent quality.
Good solvents result in small pores, while worse solvents yield much larger pores. 
We demonstrate the control over pore structure by creating a superporous, rapidly-swelling hydrogel.

Our results provide a framework for the rational design of porous gels. 
In particular, we have shown how one can use literature polymer-solvent compatibility data to create a phase diagram that predicts phase separation (i.e. the formation of porous gels) as a function of the properties of the precursor solution.
In general, the results show that phase separation is more likely in gels with low polymer content and lower molecular weight between crosslinkers.
Interestingly, this is the opposite of `normal' phase separation in polymer solutions, where phase separation is less likely in solutions with low polymer content (\emph{e.g.} \cite{fernandez2025thermodynamics}).

There are several important directions for future research.
While our results have focused on gels formed from functional oligomers, we expect that results should also apply to gels formed from different components. 
Thus, it will be important to test our conclusions on a wide variety of different gels including hydrogels made from mixtures of mutually-reactive oligomers \cite{ishikawa2023percolation}, and chemical gels like polyacrylamide that are made from solutions containing both monomer and crosslinker molecules.
A further question is whether we can quantitatively predict the size and distribution of pores that form during phase separation. 
While we have shown that pore size is strongly affected by solvent quality, we also anticipate that it is linked to how far the polymer concentration in a gel is below the phase-separation boundary: $\phi_{ps}-\phi_p$.
Ultimately, understanding PIPS should unlock new opportunities for porous-gel applications in areas such as tissue engineering \cite{dudaryeva2024tunable,muller2025cell}, soft robotics \cite{wang2024tough}, and drug delivery \cite{annabi2010controlling, tokarev2010stimuli}, and give insights into phase separation processes in related fields (\emph{e.g.} biophysical phase separation inside living cells \cite{mittag2022conceptual}). 

\section*{Methods}

\subsection*{Gel preparation}
To prepare PEGDA precursor solutions, we mix polyethylene glycol diacrylate (PEGDA) with molecular weights of 575 and 700~g/mol (Sigma-Aldrich) with various solvents, including deionized water (18.2~M$\Omega\cdot$cm, Milli-Q), ethanol (Merck, absolute for analysis), isopropanol (Sigma-Aldrich, $\geq 99.8\%$), \emph{sec}-butanol (Thermo Scientific, $\geq 99\%$), acetone (Sigma-Aldrich, $\geq 99.5\%$), and acetonitrile (Sigma-Aldrich), at polymer concentrations ranging from 10 to 90~vol\%. 
All chemicals are used as received.
A photoinitiator, 2-hydroxy-2-methylpropiophenone (Tokyo Chemical Industry), is added at 0.1~vol\% of the final volume. 
The precursor solutions are thoroughly mixed and sonicated for 10 minutes to ensure homogeneity. 
We then transfer the solutions into 2-mm-thick quartz cuvettes and seal them with parafilm to prevent evaporation. 
The cuvettes are exposed to UV light (wavelength: 365~nm; power density: 10~mW/cm$^2$~$\pm$~1~mW/cm$^2$) for 1 hour to induce crosslinking. 
The resulting gels are subsequently analyzed for their optical properties.

\subsection*{Cryo-EM}
We prepare samples for cryo-electron microscopy by plunge-freezing them manually in a liquefied ethane/propan mixture (37\%/63\%, Carbagas), and storing them in custom-made aluminum boxes under liquid nitrogen (LN\textsubscript{2}). 
Prior to imaging, we load the samples under LN\textsubscript{2} onto a cryo-table (Bal-Tec) inside a glovebox (MicroscopySolutions). 
We transfer the cryo-table via a VCT-100 cryo-vacuum shuttle (Bal-Tec) into a pre-cooled sample preparation tool (BAF-060, Bal-Tec) at --120$^\circ$C. Samples are freeze-etched at --110$^\circ$C for 1 minute, then coated at --120$^\circ$C with tungsten via electron evaporation: 3~nm at a 45$^\circ$ angle and 3~nm from the top (90$^\circ$). 
After coating, we transfer the samples via VCT-100 to a Zeiss Leo-1530 SEM equipped with a Bal-Tec cryo system and image them at --120$^\circ$C using the Everhart-Thornley and/or in-lens detector at 2~kV acceleration voltage.

\subsection*{Transmission measurements}
We characterize the optical properties of precursor solutions and crosslinked gels using a UV-Vis spectrophotometer (Cary 60, Agilent Technologies). 
Transmission spectra are recorded in the wavelength range of 200–800~nm with a resolution of 5~nm in transmission mode. 
All measurements are performed directly in the quartz cuvettes used for gel formation. 
We analyze the data in MATLAB to determine transmission as a function of polymer concentration. Transmission values (in percent) are converted to absorbance using the relation:
\begin{equation}
A = -\log_{10} \left(\frac{T_{\%}}{100} \right)
\end{equation}

\subsection*{Swelling Measurements}
To quantify gel swelling, we prepare PEGDA precursor solutions following the procedure described above. 
For ease of demolding, we use a different mold: a 2-mm-thick spacer is adhered to a glass plate, and another removable glass plate is placed on top. 
After curing, we remove the samples, gently pat them with lint-free tissue, and measure their initial mass. 
The samples are then immersed in water baths, and their mass is monitored periodically until equilibrium swelling is reached—defined as the point at which no further mass change is observed.

\subsection*{Microrheology}
We perform microrheology to measure the viscosity of PEGDA solutions at various polymer concentrations. 
Fluorescent nanoparticles (500~nm diameter, green, carboxylate-modified Fluospheres, Thermo Fisher Scientific) are added to solutions at 0.2~vol\% as tracer particles.
Imaging chambers are assembled by placing SecureSeal spacers (Grace Bio-labs, 9~mm diameter, 120~$\mu$m depth) onto glass slides and sealing them with a second glass slide after introducing the polymer solution.
We track the thermal motion of particles in 2D using a Nikon Ti2 Eclipse confocal microscope with a spinning disk unit (3i) and a 488~nm laser. 
Images are acquired using a 60$\times$ water-immersion objective (NA 1.2) with a 15~ms exposure time and 45~ms time interval. We perform particle tracking using MATLAB, and calculate the viscosity from measurements of particle diffusivity, following \cite{vestergaard2014optimal}.
The reported viscosity values represent the average of at least 5000 total observations to ensure statistical significance. 
For each sample, we acquire images at three different locations within the imaging chamber to assess variability and determine error bars.

To estimate the overlap concentration, we use an established relationship between intrinsic viscosity and $\phi^*$~\cite{rodrigues2020critical}:
\begin{equation}
\phi^* = \frac{1.45}{[\eta]}
\end{equation}
Intrinsic viscosity $[\eta]$ is estimated from the Taylor expansion of the relative viscosity $\eta/\eta_0$:
\begin{equation}
\frac{\eta}{\eta_0} = 1 + [\eta]\phi + k[\eta]^2\phi^2 + \dots
\end{equation}
Retaining only the linear term and substituting $\phi = \phi^*$ yields:
\begin{equation}
\eta = \eta_0 \left(1 + \frac{1.45}{\phi^*} \phi \right)
\end{equation}

\subsection*{PEG solubility check}
We prepare solutions of polyethylene glycol (PEG) with a molecular weight of 12,000~g/mol (Sigma-Aldrich) by dissolving the polymer in various solvents, including deionized water (18.2~M$\Omega\cdot$cm, Milli-Q), ethanol (Merck), isopropanol (Sigma-Aldrich), \emph{sec}-butanol (Thermo Scientific), acetone (Sigma-Aldrich), and acetonitrile (Sigma-Aldrich), at a polymer concentration of 10~vol\%. 
The mixtures are placed in glass vials and subjected to shaking, vortexing, and sonication to facilitate dissolution. 
Mixing continues until the solutions appear either homogeneous or no visible big aggregates or clumps remain.

\subsection*{Swelling kinetics of superporous gels}
PEGDA with a molecular weight of 700~g/mol is mixed with deionized water and isopropanol at 20~vol\% concentration, respectively. 
The precursor solutions are prepared following the same procedure described previously, then poured into molds consisting of two glass plates separated by a 5~mm spacer and cured under UV light. 
The resulting hydrogels are carefully removed from the molds, gently blotted to remove surface moisture, and weighed. 
The gels are then placed in Petri dishes and left to dry under ambient conditions until the weight stabilizes. 
To ensure further drying, the samples are transferred to a vacuum oven at 60$^\circ$C. 
Drying continues until the polymer content exceeds 85~wt\%. 

After drying, the samples are glued side by side onto a vertical-oriented, transparent plastic plate using adhesive.
The samples are then dipped into a dilute solution of rhodamine dye in water, and the sample swelling is then recorded via a video capture.

\begin{acknowledgments}

We thank Se-Hyeong Jung, Alba Sicher, Kathryn Rosowski, Lucien Cousin and Mark Tibbit for helpful discussions. We thank ScopeM and Lucas Falk  for cryo-electron imaging. 
We acknowledge support from the Swiss National Science Foundation (200021-212066).

\end{acknowledgments}

\appendix


%

\end{document}